\documentclass[aps,prl,twocolumn,superscriptaddress, showpacs]{revtex4}

\usepackage{graphicx}
\usepackage{longtable}

\newcommand{\peppo}{\(p \rightarrow e^+ \pi^0\)}
\newcommand{\pmppo}{\(p \rightarrow \mu^+ \pi^0\)}

\begin{document}


\title{Search for Proton Decay via $p \to e^+\pi^0$ and $p
\to\mu^+\pi^0$
in a Large Water Cherenkov Detector} 

\newcommand{\AFFicrr}{\affiliation{Kamioka Observatory, Institute for Cosmic Ray Research, University of Tokyo, Kamioka, Gifu 506-1205, Japan}}
\newcommand{\AFFkashiwa}{\affiliation{Research Center for Cosmic Neutrinos, Institute for Cosmic Ray Research, University of Tokyo, Kashiwa, Chiba 277-8582, Japan}}
\newcommand{\AFFipmu}{\affiliation{Institute for the Physics and
Mathematics of the Universe, University of Tokyo, Kashiwa, Chiba
277-8582, Japan}}
\newcommand{\AFFbu}{\affiliation{Department of Physics, Boston University, Boston, MA 02215, USA}}
\newcommand{\AFFbnl}{\affiliation{Physics Department, Brookhaven National Laboratory, Upton, NY 11973, USA}}
\newcommand{\AFFucd}{\affiliation{Department of Physics, University of California, Davis, Davis, CA 95616, USA}}
\newcommand{\AFFuci}{\affiliation{Department of Physics and Astronomy, University of California, Irvine, Irvine, CA 92697-4575, USA }}
\newcommand{\AFFcsu}{\affiliation{Department of Physics, California State University, Dominguez Hills, Carson, CA 90747, USA}}
\newcommand{\AFFcnm}{\affiliation{Department of Physics, Chonnam National University, Kwangju 500-757, Korea}}
\newcommand{\AFFduke}{\affiliation{Department of Physics, Duke University, Durham NC 27708, USA}}
\newcommand{\AFFgmu}{\affiliation{Department of Physics, George Mason University, Fairfax, VA 22030, USA }}
\newcommand{\AFFgifu}{\affiliation{Department of Physics, Gifu University, Gifu, Gifu 501-1193, Japan}}
\newcommand{\AFFuh}{\affiliation{Department of Physics and Astronomy, University of Hawaii, Honolulu, HI 96822, USA}}
\newcommand{\AFFkanagawa}{\affiliation{Physics Division, Department of Engineering, Kanagawa University, Kanagawa, Yokohama 221-8686, Japan}}
\newcommand{\AFFkek}{\affiliation{High Energy Accelerator Research Organization (KEK), Tsukuba, Ibaraki 305-0801, Japan }}
\newcommand{\AFFkobe}{\affiliation{Department of Physics, Kobe University, Kobe, Hyogo 657-8501, Japan}}
\newcommand{\AFFkyoto}{\affiliation{Department of Physics, Kyoto University, Kyoto, Kyoto 606-8502, Japan}}
\newcommand{\AFFumd}{\affiliation{Department of Physics, University of Maryland, College Park, MD 20742, USA }}
\newcommand{\AFFmit}{\affiliation{Department of Physics, Massachusetts Institute of Technology, Cambridge, MA 02139, USA}}
\newcommand{\AFFmiyagi}{\affiliation{Department of Physics, Miyagi University of Education, Sendai, Miyagi 980-0845, Japan}}
\newcommand{\AFFnagoya}{\affiliation{Solar Terrestrial Environment
Laboratory, Nagoya University, Nagoya, Aichi 464-8602, Japan}}
\newcommand{\AFFsuny}{\affiliation{Department of Physics and Astronomy, State University of New York, Stony Brook, NY 11794-3800, USA}}
\newcommand{\AFFniigata}{\affiliation{Department of Physics, Niigata University, Niigata, Niigata 950-2181, Japan }}
\newcommand{\AFFokayama}{\affiliation{Department of Physics, Okayama University, Okayama, Okayama 700-8530, Japan }}
\newcommand{\AFFosaka}{\affiliation{Department of Physics, Osaka University, Toyonaka, Osaka 560-0043, Japan}}
\newcommand{\AFFseoul}{\affiliation{Department of Physics, Seoul National University, Seoul 151-742, Korea}}
\newcommand{\AFFshizuokasc}{\affiliation{Department of Informatics in
Social Welfare, Shizuoka University of Welfare, Yaizu, Shizuoka, 425-8611, Japan}}
\newcommand{\AFFshizuoka}{\affiliation{Department of Systems Engineering, Shizuoka University, Hamamatsu, Shizuoka 432-8561, Japan}}
\newcommand{\AFFskk}{\affiliation{Department of Physics, Sungkyunkwan University, Suwon 440-746, Korea}}
\newcommand{\AFFtohoku}{\affiliation{Research Center for Neutrino Science, Tohoku University, Sendai, Miyagi 980-8578, Japan}}
\newcommand{\AFFtokyo}{\affiliation{The University of Tokyo, Bunkyo, Tokyo 113-0033, Japan }}
\newcommand{\AFFtokai}{\affiliation{Department of Physics, Tokai University, Hiratsuka, Kanagawa 259-1292, Japan}}
\newcommand{\AFFtit}{\affiliation{Department of Physics, Tokyo Institute
for Technology, Meguro, Tokyo 152-8551, Japan }}
\newcommand{\AFFtsinghua}{\affiliation{Department of Engineering Physics, Tsinghua University, Beijing, 100084, China}}
\newcommand{\AFFwarsaw}{\affiliation{Institute of Experimental Physics, Warsaw University, 00-681 Warsaw, Poland }}
\newcommand{\AFFuw}{\affiliation{Department of Physics, University of Washington, Seattle, WA 98195-1560, USA}}

\AFFicrr
\AFFkashiwa
\AFFipmu
\AFFbu
\AFFbnl
\AFFucd
\AFFuci
\AFFcsu
\AFFcnm
\AFFduke
\AFFgifu
\AFFuh
\AFFkanagawa
\AFFkek
\AFFkobe
\AFFkyoto
\AFFmiyagi
\AFFnagoya
\AFFsuny
\AFFniigata
\AFFokayama
\AFFosaka
\AFFseoul
\AFFshizuoka
\AFFshizuokasc
\AFFskk
\AFFtokai
\AFFtokyo
\AFFtsinghua
\AFFwarsaw
\AFFuw

\author{H.~Nishino} 
\AFFkashiwa
\author{S.~Clark} 
\AFFbu

\author{K.~Abe}
\AFFicrr
\author{Y.~Hayato}
\AFFicrr
\AFFipmu
\author{T.~Iida}
\author{M.~Ikeda}
\author{J.~Kameda}
\author{K.~Kobayashi}
\author{Y.~Koshio}
\author{M.~Miura} 
\AFFicrr
% ...

\collaboration{The Super-Kamiokande Collaboration}
\noaffiliation

\date{\today}

\begin{abstract}

We have searched for proton decays via \peppo\  and \pmppo\ using data from
a 91.7 kiloton$\cdot$year exposure of Super-Kamiokande-I and a 49.2
kiloton$\cdot$year exposure of Super-Kamiokande-II.
No candidate events were observed with expected backgrounds induced by
atmospheric neutrinos of 0.3 events for each decay mode.
From these results, we set lower limits on the
partial lifetime of 
$8.2\times10^{33}$ and $6.6\times10^{33}$ years 
 at 90\% confidence level
for \peppo\ and \pmppo\ modes, respectively.

\end{abstract}


\maketitle

One of the unique predictions of Grand Unified Theories (GUTs) is baryon
number violation. Via the exchange of a very heavy gauge boson, two
quarks in a proton can transform into a lepton and an anti-quark
resulting in a lepton plus meson final state. Because the mass of the
gauge boson is predicted to be very heavy, such processes are expected
to be very rare. The prototypical GUT, minimal SU(5)\cite{Georgi:1974sy,
Langacker:1980js, Langacker:1994vf},
predicts the lifetime of the proton when it decays to $e^+ \pi^0$ to
be less than $10^{32}$ years. A number of other GUT models
favor proton decay into this final state\cite{Lee:1994vp,Covi:2005pd},
 or with comparable branching into the similar final state
$\mu^+ \pi^0$\cite{Ellis:2002vk,Barr:1981qv}. Contemporary theories must
take steps
to evade the stringent limits set by Super-Kamiokande\cite{Shiozawa:1998si}
and prior experiments\cite{McGrew:1999nd, Hirata:1989kn} that decisively ruled out minimal
SU(5). 
One means is by introducing supersymmetry, which raises the
mass of the heavy gauge boson, but which introduces other decay
channels which are also constrained by experiment\cite{Kobayashi:2005pe}.

In this paper, we describe our search for proton decay by the reactions
$p \rightarrow e^+\pi^0$ and $p \rightarrow \mu^+\pi^0$. The result of
the search was negative, no proton decay was found. We update our
lifetime limit on the $e^+\pi^0$ channel based on six times greater
exposure than our previous publication\cite{Shiozawa:1998si}; we report for
the first time our limit on the $\mu^+ \pi^0$ channel.  Approximately
1/3 of the exposure was with photocollection coverage reduced by a
factor of two, so we also describe the effect of reduced photocoverage
on these studies.

Super-Kamiokande is a 50-kiloton water Cherenkov detector located in the
Kamioka Observatory in Japan\cite{Fukuda:2002uc}. 
The 22.5-kiloton fiducial volume of the detector was viewed by 11146 20-inch diameter
photomultiplier tubes (PMTs).
Super-Kamiokande started observation in April 1996 and
stopped in September 2001 (SK-I) for a detector upgrade.
On November 12, 2001, during water filling after the upgrade, 
a shock wave was initiated by an imploding bottom PMT. 
About half of the PMTs were lost in this accident.
In the first recovery stage (SK-II), 5182 PMTs were enclosed in a fiber 
reinforced plastic case with an acrylic cover for protection against
another chain reaction.
SK-II re-started observation in October 2002 and stopped in October 2005
for full detector reconstruction. 
The transparency and the reflection of the acrylic cover were 97\% and 1\%, 
respectively. 
The photo-sensor coverage was 40\% and 19\% in SK-I and SK-II,
respectively.

A Monte Carlo simulation was used to estimate the efficiency of detecting
proton decay occurring in water ($\mathrm{H_{2}O}$). The two free protons and
eight bound protons in an $\mathrm{H_{2}O}$ molecule are assumed to decay
with equal probability.
For the case of a free proton in hydrogen, the momenta of the decay particles
are uniquely determined by two-body kinematics. For the case of a bound
proton in oxygen, the decay particle momenta are no longer determined by
simple two-body decay; this is due to Fermi motion of the protons and the
nuclear binding energy, as well as meson-nuclear interactions in oxygen.

The Fermi momentum distribution of the nucleons was taken to be the same as
that measured by electron scattering on $\mathrm{^{12}C}$ for S-states
and P-states\cite{Nakamura:1976mb}. Nuclear binding energy was
taken into account by using a modified proton mass given by $M'_{p} =
M_{p} - E_{B}$, where $M_{p}$ is the proton rest mass, and $E_{B}$ is the
nuclear binding energy. The value of $E_{B}$ for each simulated event was
randomly selected from a Gaussian probability density function with
$(\mu,\sigma) = (39.0,10.2)$~MeV for the S-state and $(\mu,\sigma) =
(15.5,3.82)$~MeV for the P-state.
Pions interact strongly in the nucleus and may undergo scattering, charge
exchange, or absorption as they travel through the nucleus. These
interactions affect our ability to reconstruct the $p \rightarrow e^{+}
\pi^{0}$ event, and were carefully simulated to give our best estimate
of the detection efficiency of our expected signal. 
The cross sections used in the simulation of each type
of pion-nucleon interaction were calculated by the model of Oset {\it{et
al.}} \cite{Salcedo:1987md}.
Another effect in the nuclear medium which we take to occur
for approximately $10\%$ of
proton decay events is that of correlated decay: $pN \rightarrow
e^{+}\pi^{0}N$~\cite{Yamazaki:1999gz}.
The
value of the invariant mass calculated in the case of correlated decays is
lower than that of standard (uncorrelated) proton decays, leading to a
reduction in the proton decay detection efficiency when an invariant mass
selection criterion is applied.

Backgrounds to the proton decay search arise from atmospheric neutrino
interactions which may mimic the $p \rightarrow e^{+}\pi^{0}$ and
$p \rightarrow \mu^{+}\pi^{0}$ signals by charged current interactions such
as $\nu N \rightarrow \ell N' \pi^{0}$ and neutral current interactions such
as $\nu N \rightarrow \nu N' \pi (\pi '\textrm{s})$. Backgrounds were estimated using the
{\tt{neut}}\cite{Hayato:2002sd} neutrino interaction Monte Carlo
simulation with an
input atmospheric neutrino flux\cite{Honda:2006qj}.

Particles produced in the simulation of proton decay
events and atmospheric neutrino interactions were passed through a
GEANT-3 \cite{geant} based custom detector simulation to model 
Cherenkov light emission from charged particles,
particle and light propagation through matter, 
detector geometry, 
and the response of the PMTs and data acquisition electronics. 
Hadronic interactions were treated by CALOR \cite{Gabriel:1989ri}
for nucleons and charged pions of $p_{\pi} > 500~$ MeV/$c$,
and by a custom simulation program \cite{Nakahata:1986zp} for charged
pions of $p_{\pi} \leq 500~$ MeV/$c$.

We used data from 
a 91.7 kiloton$\cdot$year exposure of 1489 live days during SK-I 
and a 49.2 kiloton$\cdot$year exposure of 798 live days during SK-II.
There were about
$10^6$ event triggers/day above a threshold of a few MeV. We applied
several stages of data reduction in order to remove background from
cosmic rays, flashing PMTs and 
radioactivities.
In SK-I and SK-II, 12232 and 6584 events were obtained, respectively.
The background
contamination of non-neutrino interactions in the final
fiducial volume sample was negligible, less than 1\%. 
The 
efficiency for proton decay events to appear in this sample was
estimated to be greater than 99\%.

For events after the chain of reduction processes, physical
quantities were reconstructed using timing and charge information from
each PMT. Software reconstruction algorithms are shared between
SK-I and SK-II analyses, taking into account the
different PMT densities. The
directions and opening angles of visible Cherenkov rings were
reconstructed with a likelihood method, using the fitted vertex
position where the time-of-flight subtracted timing distribution was
most sharply peaked.  Each identified Cherenkov ring was categorized as
a showering electron-like (\(e\)-like) or a non-showering muon-like
(\(\mu\)-like) ring.  Momentum was determined by the sum of
photoelectrons corrected for light attenuation in water, PMT angular
acceptance, and PMT coverage. 
Finally, the number of electrons from muon decay
was determined by searching for delayed electron signals.
The delayed signals were found as delayed timing peaks within 
the primary event (up to 800~ns from the trigger time),
or as subsequent event triggers within 20~$\mu$s after the primary event.
In multi-ring events such as proton decay signals, each PMT observes
an integrated sum of Cherenkov light coming from all rings.
In order to determine the momentum for each ring, the fraction of the
photoelectrons from each ring was estimated for each PMT using
an expected Cherenkov light distribution for each ring.
Those photoelectrons were corrected by 
the amount of
scattered Cherenkov light in water, reflected light on PMT surfaces, and
the vertex position shift due to the \(\gamma\)'s conversion length.

For SK-I (SK-II) free proton decays of \peppo{},
the estimated vertex resolution was 18.1 (20.1) cm, the fraction of 2-ring
and 3-ring
events was 39$\pm2$\% and 60$\pm2$\% (38$\pm2$\% and 60$\pm2$\%), and the
$\mu/e$ particle misidentification
probability was estimated to be 3.3\% (3.4\%).
We define the total momentum in an event as
\(P_{tot} = |\sum_{i}^{all~rings} \vec p_i|\), where \(\vec p_i\)
is reconstructed momentum vector of the \(i\)-th ring.  
The total invariant mass
is also defined as \(M_{tot} = \sqrt{E_{tot}^2 - P_{tot}^2}\), where
total energy \(E_{tot} = \sum_{i}^{all~rings} \sqrt{|\vec p_i|^2 +
m_i^2}\),
and $m_i$ is assumed to be the electron (muon) rest mass for $e$-like
($\mu$-like) ring.
The total invariant mass 
for simulated free proton decays
was well reconstructed both in SK-I and SK-II.
The mean of reconstructed masses agreed within 1\% between SK-I and SK-II.
The resolution of the total momentum and total invariant mass
was 30.5 MeV/\(c\) and 28.7 MeV/\(c^2\) for SK-I,
and 36.6 MeV/\(c\) and 38.4 MeV/\(c^2\) for SK-II, respectively.


The energy scale stability of the detector was tested by the
mean reconstructed energy of stopping cosmic ray muons and
their decay electrons.  It varied within \(\pm 0.88\)\% (\(\pm 0.55\)\%)
over the SK-I (SK-II) period.
The absolute energy scales for SK-I and SK-II
were separately adjusted by using observed photo electrons of 
penetrating cosmic ray muons in each data taking period. 
The scale was checked by many calibration data
such as stopping cosmic ray muons, 
electrons from their decays,
and the invariant mass of $\pi^0$s produced in atmospheric 
neutrino interactions.
From comparisons of these sources
and MC simulation, the absolute calibration error was estimated to be less
than \(\pm 0.74\%\) (\(\pm 1.6\%\)) for SK-I (SK-II) period.

  In order to select proton decay signals from the data samples, 
  the following selection criteria were applied:
  (A) The number of rings is two or three.
  (B) One of the rings is $e$-like ($\mu$-like) for \peppo\ (\pmppo) and
  all the other rings are $e$-like.
  (C) For three ring events, 
  $\pi^0$ invariant mass is reconstructed between 85 and 185~MeV/$c^2$.
  (D) The number of electrons from muon decay is 0 (1) for \peppo\ (\pmppo).
  (E) The reconstructed total momentum is less than 250 MeV/$c$,
  and the reconstructed total invariant mass is between 800 and 1050~MeV/$c^2$.

  
  The detection efficiencies at each selection criterion for the \peppo\
  and
  \pmppo\ search
  are shown in Table~\ref{table_efficiency}.
  The efficiencies were
  estimated to be
  44.6\% (43.5\%) and 35.5\% (34.7\%) for SK-I (SK-II), respectively.
  The difference between the detection efficiencies of SK-I and SK-II
  was only 1\%, which implies that
  SK-II has a similar performance with SK-I for these types of proton
  decay events.
  The inefficiency for proton decay detection
  is mainly due to nuclear interaction effects of pions in $^{16}$O.
  In the simulated proton decay samples, 
  about 37\% of $\pi^0$'s
  from proton decay in $^{16}$O were absorbed or
  charge-exchanged by interactions with nucleons.
  Those events rarely survive the selection criteria.
  The efficiency for \pmppo\ was lower than that for \peppo\
  because in criterion (D)
  one electron from muon decay was required for the \pmppo\ search
  and the detection efficiency for electrons from muon decay
  was approximately 80\%.

 \begin{table}
  \caption{Number of surviving events in the data and the atmospheric
  neutrino MC 
  at each selection criterion in 91.7(SK-I)+49.2(SK-II) kiloton$\cdot$year exposure.
  Detection efficiencies using the  proton decay MCs are also listed.
  \label{table_efficiency}}
  \begin{ruledtabular}
   \begin{tabular}{lcc|c}
    Criteria & \peppo & \pmppo & Efficiency(\%)\\
    & data (atm.MC) & data (atm.MC) & ($e^+\pi^0$) ($\mu^+\pi^0$)\\
    \hline


    
    in fiducial & 18816 (19269)  & 18816 (19269) & 98.6 \ 98.8 \\
    (A)      & 4889  (5124)   &  4889 (5124)     & 73.1 \ 73.6 \\
    (B)      & 3036  (3141)   &  1536 (1604)     & 64.7 \ 61.4 \\
    (C)      & 2541  (2613)   &  1281 (1284)     & 62.6 \ 59.7 \\
    (D)      & 1859  (1941)   &  642  (580)      & 61.8 \ 46.3 \\
    (E)      &    0  (0.3)    &    0  (0.3)      & 44.2 \ 35.3


   \end{tabular}
  \end{ruledtabular}
 \end{table}

  The number of remaining events of the atmospheric neutrino MC and the data
  at each selection criterion are also shown in Table~\ref{table_efficiency}.
  The atmospheric neutrino MC,
  in which
  neutrino oscillation was accounted for with
  $\Delta
  m^2=2.5\times10^{-3} $eV$^2$ and $\sin^22\theta=1.0$,
  was normalized with the observed data using the number of
  single-ring $e$-like events, which are assumed to have
  negligible neutrino oscillation.
  Comparison of the number of events in Table~\ref{table_efficiency}
  demonstrates that
  the atmospheric neutrino MC reproduces the observed data well.
  The number of background events from atmospheric neutrinos 
  in the 91.7+49.2 kiloton$\cdot$year exposure
  for \peppo\ and \pmppo\ selection criteria
  were estimated to be 
  $0.30\pm0.04(\textrm{MC stat.})\pm0.11(\textrm{sys.})$ events 
  ($=2.1\pm0.3\pm0.8$ events/megaton$\cdot$year)
  and 
  $0.34\pm0.05(\textrm{MC stat.})\pm0.12(\textrm{sys.})$ events
  ($=2.4\pm0.4\pm0.9$ events/megaton$\cdot$year),
  respectively.
  Approximately 81\%(96\%) of background events for \peppo\ 
  (\pmppo) were due to charged
  current neutrino interactions, which consisted of 32\%(47\%) single-pion
  production,
  19\%(21\%) multi-pion production, and 28\%(15\%) quasi-elastic scattering
  (CCQE).
  In most background events from CCQE,
  a highly energetic proton ($>$1GeV/$c$) produced by the neutrino interaction
  scatters in water and produces a secondary $\pi^0$ which makes a
  similar event signature to proton decay.
  For the systematic error of the background rate,
  uncertainties of event reconstruction,
  hadron propagation in water, pion-nuclear effects, cross
  sections of neutrino interactions and neutrino fluxes were considered.

  The background event rates were also estimated by another neutrino
  interaction model, {\tt NUANCE}\cite{Casper:2002sd}, and the experimental
  data of K2K $\nu$ beam and the 1-kiloton water Cherenkov
  detector\cite{Mine:2008rt}.
  The expected background rates were $1.9\pm0.7(\textrm{MC stat.})$
  events/megaton$\cdot$year by {\tt NUANCE} for both decay modes
  and
  $1.63^{+0.42}_{-0.33}(\textrm{stat.})^{+0.45}_{-0.51}(\textrm{sys.})$
  events/megaton$\cdot$year
  by the K2K data for \peppo.
  These estimates are consistent with those from our atmospheric
  neutrino MC.

\begin{figure}
  \includegraphics[width=8.3cm]{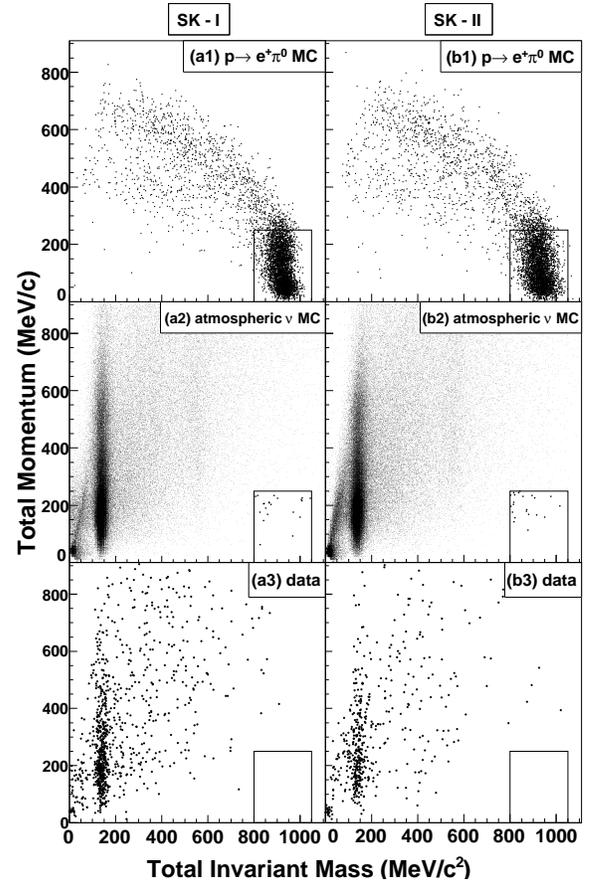}
  \caption{Total momentum versus total invariant mass distributions for
  events of
  proton decay MC (a1,b1), 500 year-equivalent atmospheric neutrino MC
  (a2,b2) and data (a3,b3) 
  which satisfy the selection criteria of \peppo\ except for (E)
  in SK-I(left figures) and SK-II(right figures).
  The boxes in figures indicates  the criterion (E).
  Points in signal boxes of the atmospheric neutrino MC (a2,b2) are
  shown in a larger size.
  }
  \label{Ptot_Mtot_epi0}
 \end{figure}

\begin{figure}
  \includegraphics[width=8.3cm]{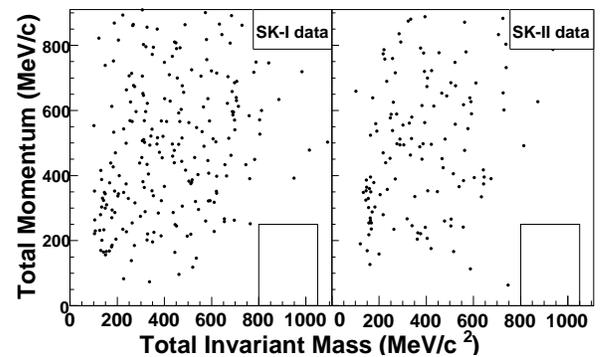}
  \caption{Total momentum versus total invariant mass distributions for
  events in SK-I(left) and SK-II(right) data, which satisfy the selection
  criteria of \pmppo\ except for (E).
  The box in figures show the criterion (E).
  }
  \label{Ptot_Mtot_mupi0}
 \end{figure}

  We have searched for proton decay signals in fully contained event
  samples
  in SK-I and SK-II.
  Figure~\ref{Ptot_Mtot_epi0}((a3) and (b3)) and Fig.~\ref{Ptot_Mtot_mupi0}
  show distributions of total
  invariant mass and
  total momentum for the samples satisfying the selection criteria except
  for (E) compared with proton decay MC (Fig.~\ref{Ptot_Mtot_epi0} (a1)
  and (b1)) and atmospheric neutrino MC (Fig.~\ref{Ptot_Mtot_epi0} (a2)
  and (b2)).
  After applying all the proton decay event selection criteria, 
  no candidate events for \peppo\ and \pmppo\ were found in the data.


  Because there were no candidates, 
  we calculated lower limits on the partial lifetime.
  In the limit calculation, uncertainties on the exposure were
  negligible.
  The total systematic uncertainty on the detection efficiency was
  estimated
  to be 19\% for both \peppo\ and \pmppo\, and for both SK-I and SK-II.
  The largest contribution to the detection efficiency uncertainty
  came from uncertainties in the
  cross sections for pion-nuclear effects in $^{16}$O; these
  were estimated to be 15\% for the \peppo\ and \pmppo\ detection
  efficiencies
  by comparing the pion escape probability from a nucleus
  with another model\cite{Bertini:1972vz}.
  The systematic error from the uncertainty of the Fermi motion was
  estimated by comparing its momentum distribution with the Fermi
  gas model;
  this changes the detection efficiencies by 9\%.
  The fraction of proton decays correlated with other nucleons in
  $^{16}$O also contributes an uncertainty that we conservatively set
   to 100\% uncertainty for a fraction of events;
  this changes the detection efficiencies by 7\%.
  There were no significant differences between
  \peppo\ and \pmppo\ in these uncertainties.
  Other sources of uncertainties from energy scale stability,
  particle identification
  and decay electron finding efficiency
  were estimated to be
  negligible.

  The lifetime limits were calculated by a method based 
  on Bayes theorem that incorporates the systematic uncertainty.
  These calculations give limits on the partial lifetime for 
  \peppo\ of $\tau/B_{p\rightarrow e^+\pi^0} > 8.2\times10^{33}$ years 
  and \pmppo\ of $\tau/B_{p\rightarrow \mu^+\pi^0} > 6.6\times10^{33}$
  years at 90\% confidence level.

We have reported the results of proton decay searches in a 140.9
kiloton$\cdot$year exposure of the Super-Kamiokande detector.
The proton decay search performance of SK-II is comparable with that of SK-I.
No evidence for proton decays via the mode \peppo\ and \pmppo\ were found
and we set limits on the partial lifetime of proton.
These limits are more stringent
compared with the previous results
of $1.6\times10^{33}$ years\cite{Shiozawa:1998si} and $4.7\times10^{32}$
years\cite{McGrew:1999nd}, respectively.

\begin{acknowledgments}

We gratefully acknowledge the cooperation of the Kamioka Mining and 
Smelting Company. The Super-Kamiokande experiment was built and has 
been operated with funding from the Japanese Ministry of Education,
Science, Sports and Culture, and the United States Department of Energy. 

\end{acknowledgments}

\bibliography{bibliography}

\end{document}